# A QoS Routing Protocol based on Available Bandwidth Estimation for wireless Ad Hoc Networks

Heni KAANICHE[1], Fatma LOUATI[2], Mounir FRIKHA[3] and Farouk KAMOUN[1]

[1] National School of Computer Sciences, Manouba, Tunisia
`Heni.KAANICHE@tunet.tn; frk.kamoun@planet.tn`

[2] National School of engineers of Sfax, Tunisia
`fatma.louati@gmail.com`

[3] Higher School of Communication of Tunis, Tunisia
m.frikha@supcom.rnu.tn

## ABSTRACT

*At the same time as the emergence of multimedia in mobile Ad hoc networks, research for the introduction of the quality of service (QoS) has received much attention. However, when designing a QoS solution, the estimation of the available resources still represents one of the main issues. This paper suggests an approach to estimate available resources on a node. This approach is based on the estimation of the busy ratio of the shared canal. We consider in our estimation the several constraints related to the Ad hoc transmission mode such as Interference phenomena. This approach is implemented on the AODV routing protocol. We call AODVwithQOS our new routing protocol. We also performed a performance evaluation by simulations using NS2 simulator. The results confirm that AODVwithQoS provides QoS support in ad hoc wireless networks with good performance and low overhead.*

## 1. INTRODUCTION

Ad hoc mobile networks, employing the IEEE 802.11 protocol in Distributed Co-ordination Function (DCF) mode, are becoming increasingly popular. In DCF mode, the 802.11 protocol does not require any centralized entity to coordinate user's transmissions. Nodes are free to move around, join and leave the network as needed. As this happens, new links form as nodes come within range of each other, and existing links break as two nodes move out of range of each other. These constant changes in topology impose a significant challenge for the communication protocols to continue to provide multi-hop communication between nodes. In fact, a key issue in MANETs is the necessity to establish an efficient and correct route between a pair of nodes so that messages may be delivered in a timely manner that what we calls the routing techniques. Several routing protocols have been developed. Such solutions must deal with the typical limitations of these networks, which include high power consumption, low bandwidth. However, most of theme considers the best effort data traffic and neglect connections with quality-of-service (QoS) requirements, such as voice channels with delay and bandwidth constraints.

Bandwidth is a crucial component of quality-of-service (QoS) in MANETs because the network topology may change constantly, and the available state information (such as the bandwidth) for routing is inherently imprecise. Recent years have seen a strong interest in techniques for estimating available bandwidth along a path in Ad hoc Networks. The available bandwidth between two neighbour nodes is defined as the maximum throughput that can be transmitted between these two peers without disrupting any ongoing flow in the network [1]. In fact, available bandwidth estimation is useful for path selection in Ad hoc networks.

Actually, our goal is to compute estimations in order to provide accurate guarantees to applications and ensuring that guarantees offered to ongoing applications in the network still hold if a new flow is accepted and shaped according to our estimation. To compute the value of this remaining bandwidth, each node uses only its local perception to evaluate the proportion of time the medium is free. This measurement can give indications on the remaining bandwidth. Our scheme does not modify the CSMA/CA MAC protocol in any manner, but gauges the effect of phenomena such as medium shared, RTS/CTS mechanism, interference, which influence the available bandwidth, on it.

We propose, in this paper, a new model to estimate available bandwidth estimation. Section 2 briefly discusses some related works in the area. We expose the constraints that effect the estimation of available bandwidth in section 3. Those constraints are obtained throw an experimental and theoretical studies. In section 4, we present the technique to calculate the foccup_bp that provides an accurate available bandwidth. We focus after that on the routing strategy that employs the foccup_bp in section 5. And Section 6 shows some experimental evaluations.

## 2. RELATED WORKS

With the applications over 802.11 WLAN increasing, the customers demand more and more new features and functions of such networks. One very important feature is the support of applications with Quality of Service (QoS) in 802.11 WLAN [17]. So, the support of video, audio, real-time voice over IP and other multimedia applications over 802.11 WLAN with QoS requirements is the key for to be successful in wireless communications. Many researchers have shown much interest in developing new medium access schemes to support QoS.

By considering that Mac layer is the key element that provides QoS support in 802.11-based wireless networks, IEEE802.11e [2] [3] is the MAC enhancements for QoS. It adds a new function called HCF (Hybrid Coordination Function). HCF supports both differentiated and parameterized QoS through prioritized contention-based and controlled contention-free medium access. QoS features of the 802.11e standard are beneficial to prioritize for example voice and video traffic over more elastic data traffic.

QoS routing in MANETS [13], as well, is an issue that has been and continues to be investigated. Recent years have seen a rush in interest in available bandwidth estimation. We propose to illustrate existing bandwidth measurement techniques for estimating available bandwidth for end-to-end paths.
Actually, available bandwidth estimation can be categorized on two major approaches: the intrusive techniques and the passive ones.

- Intrusive approaches: called active techniques as well, based on the end-to-end probe packets, needed to estimate the available bandwidth along a path. We mention DietTOPP [4] for example. It has been developed for wireless network based on the TOPP method for wired environment. His major idea is to compute the medium utilization from the delays and to derive the available bandwidth from this utilisation. The main default of such approach is the higher consummation of bandwidth.
- Passive approaches: based on the broadcast of local information on the used bandwidth via hello messages. The local information is about the channel utilization ratio, wish is deduced from a permanent monitoring of the channel status (idle or busy).

QoS-AODV [5] adapte a passive approche. It estimate the available bandwidth by defining a metric called Bandwidth Efficiency Ratio (BWER). BWER is the ratio between the number of transmitted and received packets. To collect the neighbour's available bandwidth, Hello messages are periodically broadcast in the one hop vicinity. The available bandwidth of a node is considered as being the minimum of the available bandwidth between the one hop neighbours and current node [6].

BRuIT (Bandwidth Reservation under InTerferences), a passive approach as well, takes into account the whole knowledge of interferences. In fact, BRuIT is a distributed signalling protocol which achieves this goal by periodically sending messages containing information on bandwidth availability and provides a mechanism to reserve bandwidth for transmissions. BRuIT [7] provide to the nodes information about their neighbours by broadcasting periodically hello messages. Hello packet not only includes information about the transmitter but also about every node at a distance of k hops from the transmitter. k, width of the extended neighbour hood that we consider (in other words the propagation range of the information) is a parameter of the protocol. The Hello packets are propagated within two hops.

QOLSR [8] is an enhancement of the standard OLSR. QOLSR adds extensions to the messages of control during the discovery of the neighbours. It is appropriate to insert parameters such as the delay, the band passer-by, the expense of link, loss of packet. The messages of control TC (broadcast by MPRs to announce all the knots that it not much

AQOR [9] Ad hoc Qos on-demand routing, built on top of the IEEE802.11 DCF MAC, provides end-to-end QoS support, in terms of bandwidth and end-to-end delay. This protocol, specially, offers a mechanism of reservation of bandwidth similar to a signalling protocol, the functionalities of beacon being used for the reservation of resources and path maintain. However, the proposed counting of bandwidth is complex and costly and exchanges of numerous beacons overhead the network.

## 3. VARIOUS METRICS FOR BANDWIDTH ESTIMATION

The Medium Access Control (MAC) layer uses a Carrier Sense Multiple Access with Collision Avoidance (CSMA/CA) algorithm for shared use of the medium. Unlike CSMA/CD (Carrier Sense Multiple Access/Collision Detect) which deals with transmissions after a collision has occurred, CSMA/CA acts to prevent collisions before they happen [3]. Before emitting a frame, a node senses the channel. When it is idle, the source of data have to wait a constant period of time DIFS (DCF Inter Frame Space) plus an additional random chosen time in the interval [0, contention window $_{min}$] multiplied by a time slot (20 µs) called the backoff factor so if many nodes have to emit data they can't have to wait the same time. The first node with backoff counter reaches zero can transmit the packet. Others nodes have to wait by stopping their counter and starting a deferring period until the medium becomes free again. Every succeed reception of packet must be acquitted after a constant period SIFS (Short Inter Frame Space) shorter than DIFS [10]. (Figure 1)

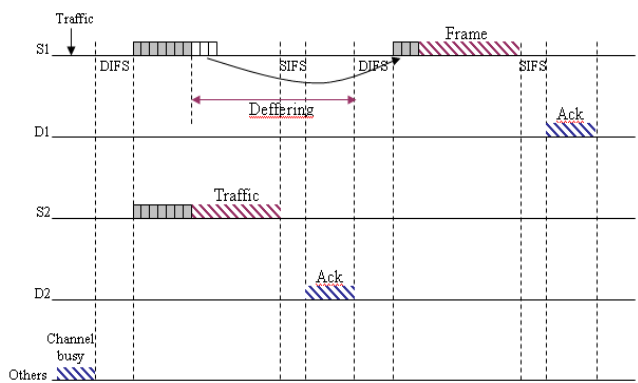

Figure 1. CSMA/CA Mechanism

### 3.1. Notation

Before introducing our QoS enhancement, let us note:
- **Basic rate:** the PLCP (Physical Layer Convergence Protocol) header of each packet is sent at the basic rate. This basic rate is 1 Mb/s for 802.11b and 6 Mb/s for 802.11a. [11]
- **Data rate:** a higher rate used to transmit the physical-layer payload (which includes the MAC header) is indicated in the PCLP header. [11]
- **Application throughput:** throughput of the data from the application layer in Open SyStem Interconnection (OSI) model.
- **Useful MAC throughput:** throughput with which useful data are sent from data link layer to physical layer.

By introducing the useful throughput versus that theoretical, we suggest that the useful throughput is less than that theoretical. Considering these results, it was imperative to us to be next to the environment of simulation NS2 to put in an obvious place these certificates. Our first accomplished simulation was aimed at determining the useful MAC throughput supported by NS2 SIMULATOR.

Our scenario consists in putting a transmitter and a distant receiver of 200 meters (the ray of the zone of communication is about 250meters). The transmitter sends in growing throughput by CBR (Constant Bit Rate) packets with 500 bytes as size. In fact, this simulation will serve us as standard for subsequent simulation. Let us say that the mobile are provided with wireless interfaces implementing the norm IEEE802.11 with bandwidth 2Mb/s. The rate simulated is about 2Mb/s since it is the one who is supported by default by NS2. Noted results are introduced in Figure2.

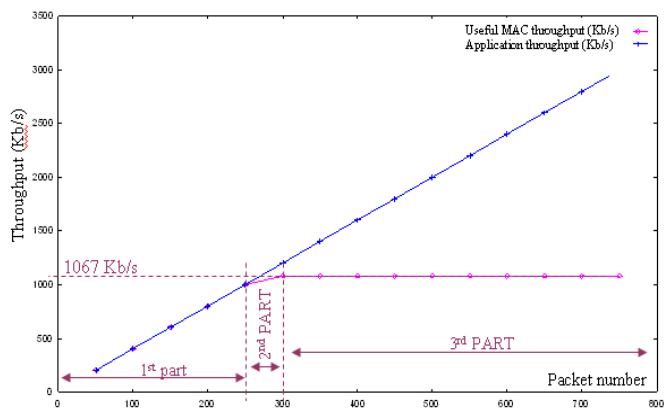

Figure 2. Useful *MAC* throughput versus application throughput (packet size 500 bytes)

Our simulation shows three parties: a first part where the useful MAC throughput follows the evolution of the application throughput. In the second part, the useful MAC throughput is less than that of the application throughput (≥1000 Kb/s). In the third part, the useful MAC throughput becomes rather constant and does not cross a value of 1067Kb/s for an application throughput equal to 300packets/s meaning superior to 1.2 Mb/s. Indeed, the useful MAC throughput assured by interfaces IEEE802.11 within NS2 can't exceed a threshold. We then undertook the same simulation unlike the size of the packets which remains 1000 bytes.

We also noted a threshold for the useful MAC throughput. However, the value of this threshold exceeds that of Figure 2. Knowing that the only corrupted parameter of simulation is the size of packets, we conclude from this that the useful MAC throughput depends on the size of the packets of the data flow being discussed. To prove our theory, we undertook the simulation of the same scenario represented above and we alter the size of packets for application throughput respective 1.2 Mb/s, 2Mb/s and 3Mb/s (Figure 3).

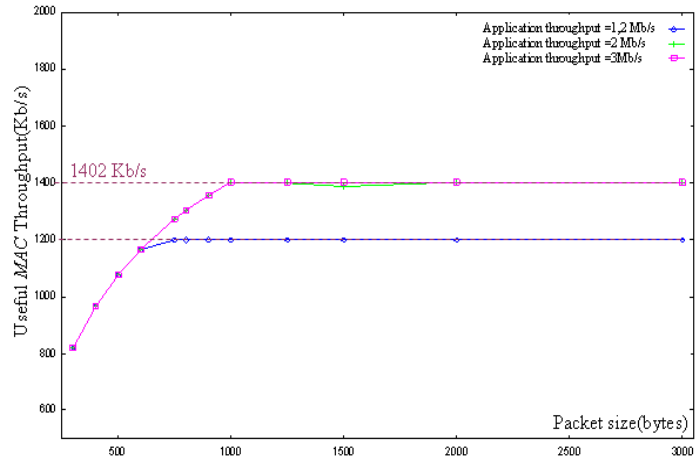

Figure 3. Evolution of the useful MAC throughput according to the size of the packets of data

### 3.2. Shared medium

The mode of wireless communication is characterized by the shared medium radio as we had already recalled it before. To demonstrate that, we considered scenario according to (Figure 4), under condition of simulation. (Table1)

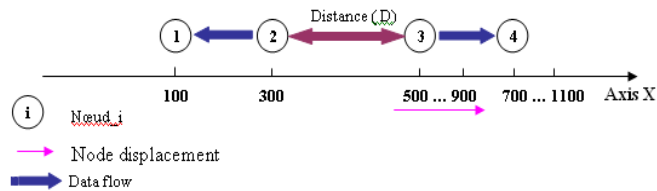

Figure 4. Shared medium (1)

Table1. Attribute values used in the simulation

| Simulator | NS2.29 | | |
|---|---|---|---|
| Topology | 1500*500 | | |
| Number of nodes | 4 | | |
| Simulation length | 300s | | |
| Routing protocol | AODV | | |
| Flow: Constant Bit Rate (CBR) | Source 2:Flow f1 | Space of time (seconds) | throughput (Kb/s) |
| | | 50s → 100s | 200 |
| | | 100s → 150s | 400 |
| | | 150s → 200s | 600 |
| | | 200s → 250s | 800 |
| | Source 3:Flow f2 | 70s → 250s | 600 |
| Packet size | 500 bytes | | |

Results are collected in Figure 5.

International Journal of Computer Networks & Communications (IJCNC) Vol.3, No.1, January 2011

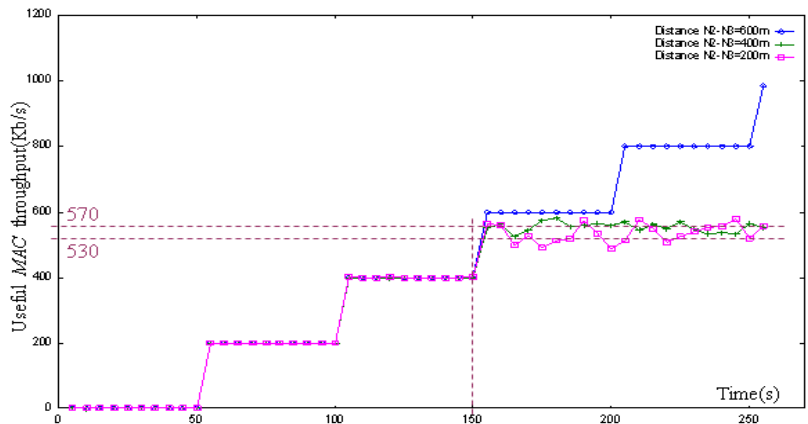

Figure 5. Evolution of the useful MAC throughput versus time at the level of pair 2-1 (1)

▪ For a distance equal to 600metters separating both communicating pairs, the useful MAC throughput follows the evolution of the application throughput imposed by scenario. In fact, the sum of both flows does not show 1400Kb/s. The useful MAC throughput assured in that case is equal to 1067Kb/s.

▪ For a distance 400metters, we clear two parts:

✓ 0-150sesonds: the sum of both flows achieves 1000Kb/s which is less than the threshold (=1067Kb/s).

✓ 150 -250seconds: the sum of flows is of 1200Kb/s during the 50 first ones seconds and on 1400 during last 50 seconds. We observe a fall of the useful MAC throughput.

There is then a specific distance for which the competitor flows shared the medium. Every source of data is going to try to make as best as he can and to send the maximum of its data packets. Given that the maximum capacity of the channel is of 1067Kb/s, the channel will be divided between these two sources.

In second time, we simulated a scenario of 250seconds in which we considered four mobile nodes among which two sources and two destinations (Figure 6). The first flow begins since instant 50seconds until instant 250seconds. While the flow2 is initialized since instant 150 second and extend until the end of scenario. Both sources are initially distant of 400metters. The results of simulation are collected in Figure 7.

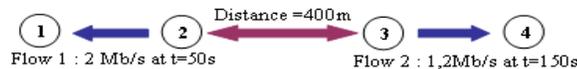

Figure 6. Shared medium (2)

The figure (7) shows a useful MAC throughput not exceeding 1400 Kb/s for a size of packets of 1000 bytes and a little more than 1000 Kb/s if their size is 500 bytes. This confirms our purposes in the previous section. In the second part of figure, the second flow comes to add up in the flow. We note a fall of the useful MAC throughput. Although the second source of the rival flow is distant of 400meters, the emission of the first source is flustered and this explains by the shared channel between the mobile nodes of network. According to our observations for both accomplished scenarios, we conclude that there is a zone beyond the zone of communication in which the emissions of one source affect those of other one. It is about the phenomenon of interference.

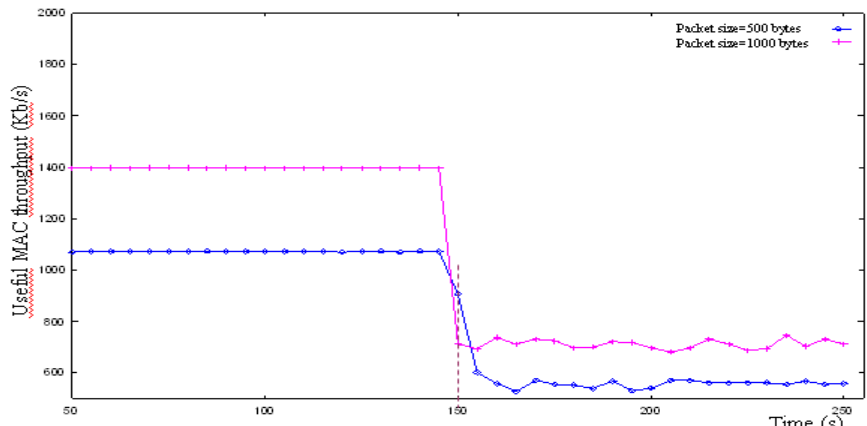

Figure 7. Evolution of the useful MAC throughput versus time at pair 2-1 (1)

### 3.3. Interferences

Interferences could decrease the applications rates. This can be a real problem for applications that need guarantees. We study, in this subsection, the impact of interferences.

At first, it was imperative to us to prove if the environment of simulation that we considered NS2 supports this phenomenon. In effect, we undertook a series of simulation in which we put two mobile nodes within reach communication. The flow of data initiated CBR is 400Kb/s. The metric that we offer to measure is PDR (Packet Delivry Ratio). The rate of issued packets PDR measures the percentage of success of the protocol. It is expressed by the number of packets of data correctly accepted by destinations wanted (Received Packet) in comparison with the packets of data issued by the sources of useful traffic (Emitted Packet).

$$PDR = \frac{\text{Received Packet}}{\text{Emitted Packet}} \times 100 \qquad (1)$$

For our scenario, PDR measured at the level of destination is 100%. We then introduced in network another communicating pair (Figure 8).

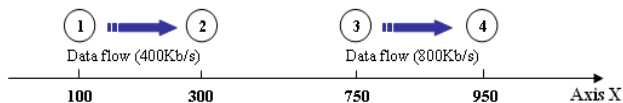

Figure 8. The phenomenon of interference under NS2 (1)

The simulation of this new network 100seconds showed that PDR measured at the level of the node 2 is equal to 61.4 %. The fall of PDR in spite of the conservation of throughput (equal to 400 Kb/s) for this pair, proves that the reception at the destination (node2) was flustered. The source of this disturbance comes necessarily from distant communicating pair of 450metters. These notes prove that NS2 takes into account the phenomenon of interference.

We started our study of this phenomenon by a second series of simulation, objective of which is to determine the ray of the zone of interference characterizing every node of network. In effect an experimental study [12] was led in this sense. Results concluded that in free space the zone of interference is nearly double the zone of communication. Our simulation (Figure 9) considers four mobile nodes divided in two communicating pairs. The parameter on which we react is the distance which separates them: Initially, both pairs are close, and then both pairs begin moving away more and more. In every pair, the transmitter and the receiver stay within the reach of communication. Traffic is a flow of CBR packets of 1000 bytes lasting 800seconds.

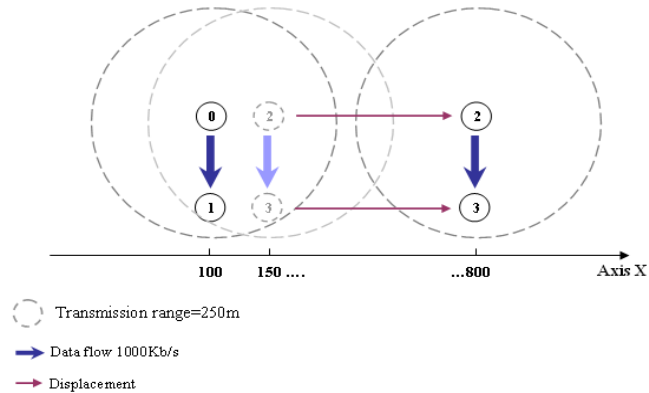

Figure 9. The phenomenon of interference under NS2 (2)

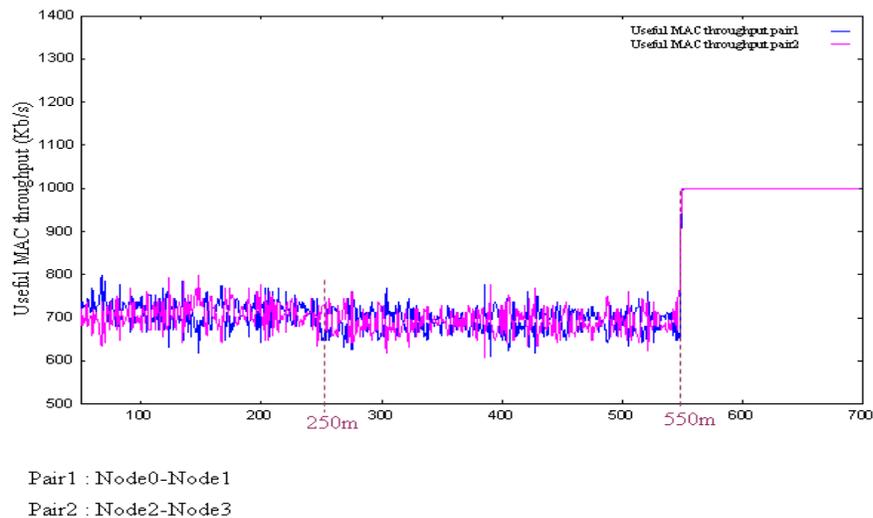

Figure 10. Useful MAC throughput versus the distance which separates both pairs

The figure 10 shows:

- For a distance ≤ 550meters: The useful MAC throughput is about 700 Kb/s at every communicating pair (less than 1000 Kb/s: application throughput of every pair). The sum of both throughputs attains 1400 Kb/s (Result already notified in the previous sections).

We also point out that for the upper distance in 250meters (range of transmission), the channel is always shared between both transmissions. This proves the existence of a zone of interference beyond the range of communication. The distribution of the channel is assured by RTS / CTS (we offer to study impact in the following subsection).

- For a distance>550meters: The useful MAC throughput in each of both pairs is 1000 Kb/s (equal to the application throughput simulated in this scenario).

Both pairs are independent, the channel is not any more shared and every source thinks it is free to emit data without being bothered therefore by the emissions of the other source.

## 4. OUR PROPOSAL

The purpose of this section is to introduce the method which we offer to estimate the factor of occupation of the medium radio in network. We define the factor of occupation of the medium as the percentage of use of the channel by traffic generated by mobile nodes being in the same zone of interference. In purpose to provide accurate available bandwidth estimation, it is necessary to be able to consider correctly the maximum throughput which every intermediate mobile node can transmit. Besides, the nature of radio between nodes in network Ad hoc provokes a new decisive point for the reliability of evaluation of the available bandwidth: the phenomenon of interferences. It is for it that our approach takes into consideration the metrics already detailed in the previous section.

The model that we offer suggests the observation of the activity of the channel radio shared, by continuous listening, to assess on a local level the occupancy rate of the medium radio and therefore the available bandwidth. Every mobile node network is capable of determining the temporal periods during which the medium is occupied and conclude as for the availability of bandwidth which is of a critical importance for QoS in Ad hoc networks.

To determine the occupation rate of the medium, every mobile node holds the temporal periods of occupation during an equal latency in Δt seconds (called observation period).

### 4.1 Determination of observation period

To determine an appropriate observation period (Δt), we considered Δt equal to 0,0025 seconds and we undertook the simulation of two mobiles according to enclosed experimental conditions (Table 2).

Table 2. Attribute values used in the simulation

| Simulator | NS2.29 |
|---|---|
| Topology | 1500*500 |
| Simulation length | 100s |
| Routing protocol | AODV |
| throughput | 5 packets/s |
| Packet size | 500 bytes |

We noticed during this simulation that for every space of time, separating two successive flows of data, there are two periods of time: the first period introduces the length during which the medium radio is occupied by different exchanges of packets of control between the nodes of considered network. The second period of time is a time of silence meaning time during which the medium is free. According to these notes, we could determine a problem concern the choice of the value of Δt. In fact, for a throughput of 5 packets/s (20 Kb/s), a data packet is initialized every 0,2 seconds. If they consider a Basic rate 1 Mb/s and Data rate 1Mb/s also, the period of occupation of the channel would be of 0,005536seconds, the period of silence is 0,194464seconds. Latency Δt (equal in 0,0025seconds) is broadly less than at the time of silence.

This could corrupt the estimation of the occupancy rate of the medium and bring to decisions based on wrong considerations. According to this study, we are going to choose a value of ∆t equal in 1 second. In fact, if we consider that the throughput of a flow is superior or equal to 1 packet/s, we would never have to meet a time of absolute silence in ∆t superior to 1 second. On the contrary, if throughput is less than 1 packet/s, the occupancy rate of the hanging channel ∆t will be null, what is logical since the throughput of considered traffic is very weak: a very lower throughput compared to 2 Mb/s assured by the protocol IEEE802.11. Therefore the occupancy rate of the medium is almost null and the channel will be able of transporting any traffic without any problem. According to these observations, the larger ∆t is, more stable are the measurements. However, ∆t should be small enough to take into account nodes mobility. So, we choose ∆t equal to 1 second.

### 4.2 Description of the model

The counting of the occupation factor $f_{occup\_bp}$ of the channel is accomplished according to equation (2):

$$f_{occup\_bp} = \frac{period\_channel\_busy(during\ \Delta t)}{\Delta t} \quad (2)$$

According to Figure 11, *period_channel_busy* is determinate (3)

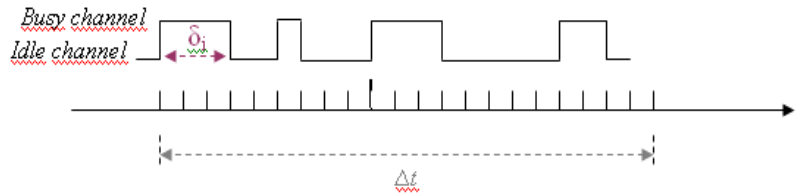

Figure 11. Busy channel

$$period\_channel\_busy = \sum_i \delta_i \quad (3)$$

To reach an estimate of the available bandwidth, it is necessary to count the traffic existing on the medium. In fact, there are three types of traffic which can influence the estimation of the $f_{occup\_bp}$ and therefore the available bandwidth for a node (I) in a network:

- *Traffic (I)*: traffic relating to the own emission of the node I.
- $Traffic_{One\_hop\_neighborhood}$ *(I)*: traffic relating to the emission of the direct neighbours of the nodes I.
- $Traffic_{Two\_hop\_neighborhood}$ *(I)*: traffic relating to the emission of the neighbours located in the zone of interferences of the mobile I.

## 5. A PROPOSED NEW PROTOCOL

In this section, we offer to introduce our new routing strategy referring to the result provided since the mechanism of evaluation of the occupancy rate of the radio medium. We have used our MAC layer bandwidth estimation scheme as an essential component in (a) Admission control and (b) reservation of resources for the construction of routing algorithm with QoS. We describe both of these in this section. Providing QoS guarantees in an Ad hoc network requires very important component admission control to ensure that the total resource requirements of admitted flows can be handled by the network. If there are not enough resources for all real time flows, some real time flows must be rejected to maintain the guarantees made to other real time flows. Our proposal is an enhancement of reactive routing protocol AODV (Ad hoc On Demand Outdistances Vector).

## 5.1 Why AODV ?

The choice of this protocol is justified according to two axes: First of all, the medium radio, shared by all mobiles, is a very rare resource for networks Ad hoc. AODV, by his reactive nature, asks for a bandwidth less important for the service of the tables of routing than the proactive protocols. In fact, these last generate massive control traffic between useful periods of communication while the reactivity of AODV reduces the load of network in term of messages of control since paths towards destinations are established only at the request of the sources of traffic of data. Besides, the periodicity of emission of the control traffic, generated by a proactive protocol, can be useless and overload network knowing that there are paths which are constructed but not used by applications.

Besides, AODV is based on the most interesting notions since DSR and DSDV such as the concept of discovery of path by Route REQuest and maintenance by Route Error of DSR as well as the principle of sequence number and the mechanism of neighbourhood discovery since DSDV. Studies in this frame revealed that the protocol AODV assures a better success (especially in term of rate of the issued packets), by comparing it with other routing protocols [14] [15]. Besides, AODV was validated as experimental RFC [16] by the working party MANET of IETF specialized in standardization of the relative clauses of routing protocols.

## 5.2 Admission control and resources reservation

Admission control is a network QoS procedure. It determines how bandwidth is allocated to stream with various requirements. In ad hoc mobile networks, Admission control is useful because of the shared medium in such kind of networks. The admission control is often coupled with a mechanism of resource reservation. In fact, any node with QoS routing protocol must be capable of reserving resources if CAC has succeed.

## 5.3 AODVwithQoS's specification

On the reception of a new application, the source (S) of a flow is going to throw a RREQ to explore the path towards destination. Any node of the network accepting this RREQ, have to make a decision: forwarding or not the RREQ. If CAC succeed then RREQ to destination is forwarding. Otherwise, RREQ is ignored. Eventually, if the node receiving RREQ is the wanted destination, then success of its control of admission must be followed by RREP (Route REPly). The migration of RREP towards the source of application is also accompanied by CAC and a real reservation of the bandwidth. The CAC during the evacuation of RREP allows making sure of the availability of requested resource. CAC that we offer demands not only the recovery of the $f_{occup\_bp}$ of the channel radio since MAC layer but also a theoretical validation of necessary time to make the monitoring of a packet of data meaning the necessary time so that a packet of data is transmitted since the source towards destination. For it, we considered two mobile nodes: a source (S) and a destination (D) (Figure 12), we are interested only in exchanges of control and data packets.

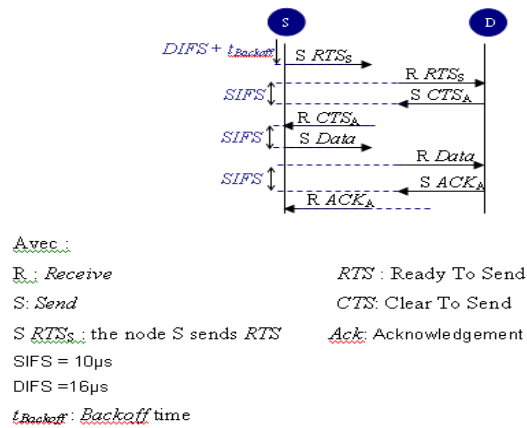

Figure 12. Flow transmission

Transmission of the traffic of data requires all its stages. To estimate the available bandwidth at the one instant of reception of a new request of resource, we offer to calculate the bandwidth that would be indeed used. From this value we can make CAC.

In fact, the counting of the bandwidth indeed requested means the counting of necessary time for the transmission of a packet. We call $t_{forwarding}$ this time (4).

$$t_{forwarding}(D) = DIFS + t_{Backoff} + t_{S\_RTS\_S} + t_{S\_CTS\_D} + t_{R\_DATA} + t_{S\_ACK\_A} + 3 \times SIFS \quad (4)$$

$t_{Backoff}$ is considered equal to the average of space multiplied by 20µs. Basic rate = 2Mb/s and Data rate= 1Mb/s.

So, $t_{forwarding}(D) \cong 0,00371 \sec onds$

We have also to recover $f_{occup\_bp}$ of the medium from the MAC layer.
We call *Bandwidth_applicative$_{available}$*(A): applicative indeed available bandwidth at the level of the node A (5)

$$Bandwidth\_applicatve_{available}(A) = \frac{(1 - f_{occup\_bp})}{t_{forwarding}} \times packet\_size_{flow} \quad (5)$$

The reception of the first request of bandwidth takes place at t=0second therefore $f_{occup\_bp}$ is null

$$Bandwidth\_applicative_{available}(D) = \frac{(1-0)}{0,00371} \times (500 \times 8) \cong 1078 Kb/s$$

Theoretical value differs lightly from that measured by simulation (=1067 Kb/ s) and this due to the fact that we considered an average of $t_{Backoff}$.

To accomplish a general implementation of counting of $t_{forwarding}$ of a packet, two problems must be treated:

The first concerns the possible membership of every node of data's path to the different zones of interferences and the second is interested to the determination of the $t_{Backoff}$.

In fact, a first resolution came to mind us: geographical approach: we envisaged that every node of network has a module of location from which, it recovers its locality since the Global Positioning System (*GPS*). Then every node will determine different interference zone that belongs to and estimate *Bandwidth_applicative$_{available}$*. In fact, the node has either to broadcast a periodical message including her location and that of her direct neighbours or to insert in *RREQ* all nodes of the path and their positions. However, we reproach to those solutions a possible

International Journal of Computer Networks & Communications (IJCNC) Vol.3, No.1, January 2011

overhead of network by signaling packets indicating location of every node or by proliferated size of the *RREQ*.

The second alternative is rather simpler. In fact, we could consider two cases:
The first case where nodes are located in different interferences zones (Figure 13)

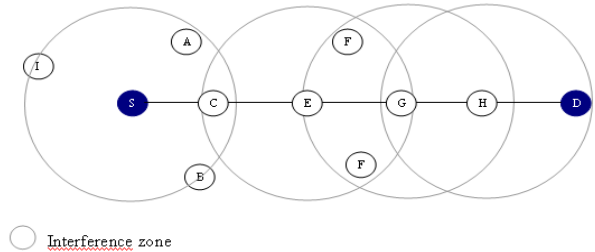

Figure 13. Overlapping of different zones of interference

The counting of the residual bandwidth of every node is going to depend on her membership to the different interferences zones. For instance the node (H) will not be bothered by the emissions of the node (A). The realization of this resolution returns to adopt geographical approach as already mentioned.

The second case to be treated is the worst case where all nodes of the path belong to the same zone of interference (Figure 14).

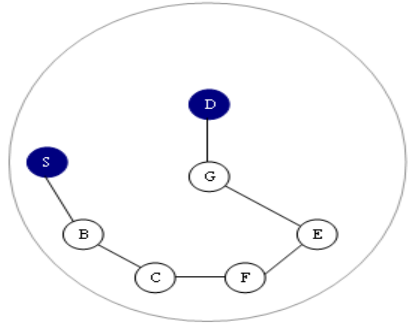

Figure 14. The worst case

In fact, for our general counting of $t_{forwarding}$, we offer to treat this case: all participants nodes in the routing of the same flow are in the same zone of interference. In fact, if we have a network with n nodes belonging to the path of data: k=n-1: meaning all nodes others than the source (S), belonging to the path, are in the same zone of interference then (6):

$$t_{forwarding}(N) = k \times (t_{RTS} + t_{CTS} + t_{ACK} + t_{DATA} + 3 \times SIFS + DIFS) + T_{Bachoff} \quad (6)$$

This general implementation raised us the second point to be treated and which is the $t_{Backoff}$. In fact, the estimation of the $t_{Backoff}$ is impossible since it is chosen randomly. So that to put right this situation, we passed by a series of simulation in which, we considered a number of variable nodes and we considered that all nodes belonging to the path of routing are in the same zone of interference. The size of the packets of the flow is 500 bytes; Basic rate is of 1Mb/s while Data rate is of 2Mb / s. Our purpose was to determine the useful MAC throughput measured according to the number of hops. Results are collected in Figure 15.

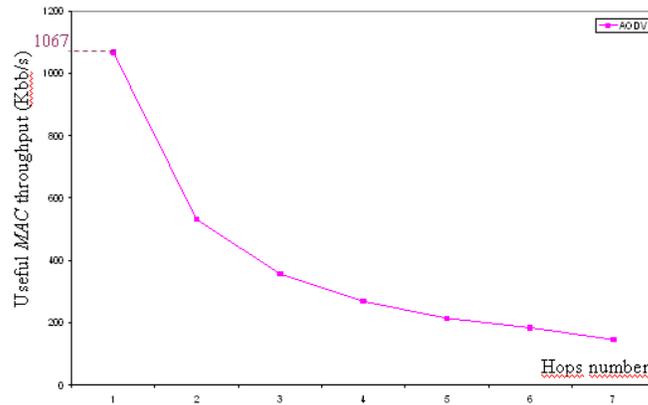

Figure 15. Useful MAC throughput versus number of hops

The speed of the curve can be assimilated with an exponential function $A \times \exp(-Bx)$ or with a hyperbolic one $\frac{A}{1+Bx}$.

We undertook the counting of these two functions and we introduced results (Figure 16).

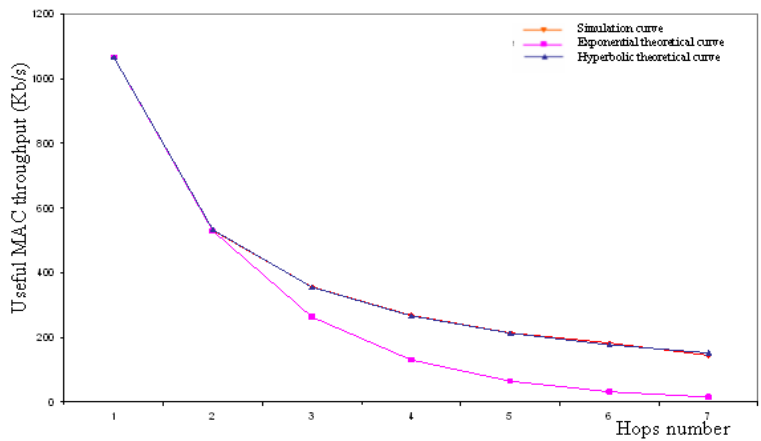

Figure 16. Approximation of the curve of the evolution of the useful MAC throughput according to the number of hops

The curve acquired by simulation match with hyperbolic function. Makes it for a total number of nodes (n) belonging to the path of a flow, we have k=n-1 number of hops (7)

$$f(n) = \frac{1067}{n-1} \Rightarrow f(k) = \frac{1067}{k} \qquad (7)$$

Now, from the estimate of the useful MAC throughput, we can deduct *Bandwidth_applicative$_{available}$* of node A by (8)

$$Bandwidth\_applicative_{available}(A) = (1 - f_{occup\_bp}) \times f(k) \qquad (8)$$

Finally, *Bandwidth_applicative $_{available}$* is calculated at each node and compared to the requested bandwidth for the flow in question by applying CAC lasting RREQ and CAC with resource reservation lasting RREP.

Let us note however that through the CAC for a flow, every node has to take into account the bandwidth already granted to other flows by considering also the average end to end delay multiplied by 2. (9)

$$Bandwidth\_applicative_{available}(N) = \qquad (9)$$

$$Bandwidth\_applicative_{available}(N) - \sum_{ins\tan t\_reception_{RREQ} \leq Ins\tan t\_reservation(i)+2\sec onds} Bande\_passante_{reserved}(i)$$

The determination of the average of end to end delay was performed by simulation. We considered rectangular ground of 1500m x 500m on which we have 50 mobiles moving according to the model Random Way Point (RWM).

The model of mobility was generated randomly by NS (Setdest), the time of repose is of 10s and we varied the speed of nodes. We indicated 20 sources of traffic of data CBR. The chosen throughput is 4 packets/s and size of packet equal to 512 bytes. Traffic CBR is generated by the generator of unpredictable traffic (Cbrgen).the routing protocol is AODV.

The average end to end delay is expressed by (10):

$$Average\ end\ to\ end\ delay = \frac{\sum_{i=1}^{Number\_received\_packet}(tr_i - te_i)}{Number\_received\_packet} \qquad (10)$$

With:
- $t_{ei}$: the instant of emission of data packet (i) by the applicative layer of the source of traffic.
- $t_{ri}$: the instant of reception of a data packet (i) by the applicative layer of the destination of traffic.c
- *Number_received_packet*: the total number of data packets accepted by applicative layers of desired destinations.

Both delays generated by the discovery of paths and delays wasted at pipes are included in the computing of average end to end delay. Data packets lost under path are not considered. The result of simulation is collected (Figure 17)

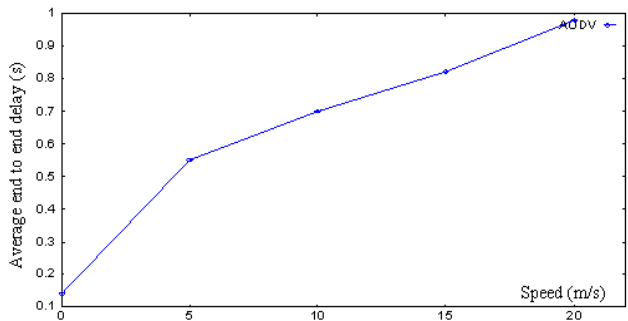

Figure 17. The average end to end delay versus speed

<a>International Journal of Computer Networks & Communications (IJCNC) Vol.3, No.1, January 2011</a>

We notice that the average delay end to end delay for a speed of 20m/s is about 1 second. Lasting the reservation, the node must take into consideration: the delay for which RREP arrives at the source and therefore in the worst cases it is equal to 1s.

The delay consumed by the first data packet relieves the path so another second. the sum makes us two seconds to be sure that by accepting a flow, they did not touch the bandwidth of those already reserved but not yet accepted by the nodes. Meaning, reserved flow is received at MAC layer and not counted in the occupation factor of the channel and so in the available bandwidth.

## 6. PERFORMANCE ANALYSIS

The purpose of our valuation is to prove that when we use our protocol AODVwithQoS, flows QoS are emitted from the source towards destination without being subjected to degradation at of their throughput. Our guarantee is the reservation mechanism that we offered. The flow requested bandwidth of which we cannot guarantee is simply rejected during the control of admission.

### 6.1 Scenario 1: simple network

The first topology which we considered is a network formed seven mobile nodes among which two sources and one destination. (Figure 18) (Table 3)

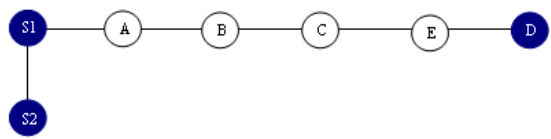

Figure 18. Static scenario

Table 3. Attribute values used in the simulation

| | |
|---|---|
| Simulator | NS2.29 |
| Topology | 1500*500 |
| Simulation length | 50s : Flow f1: de 0->50s |
| | Flow f2: de 25s->50s |
| Routing protocol | AODV |
| | AODVwithQoS |
| Traffic | CBR |
| Packet size | 500 bytes |
| Throughput | 50 packets/s=>200Kb/s |
| Transmission Range | 250m |
| Interference Range | 550m |
| RTS/CTS mechanism | Activated |
| Basic Rate | 1Mb/s |
| Data Rate | 2Mb/s |

Results are collected (Figure 19,20).

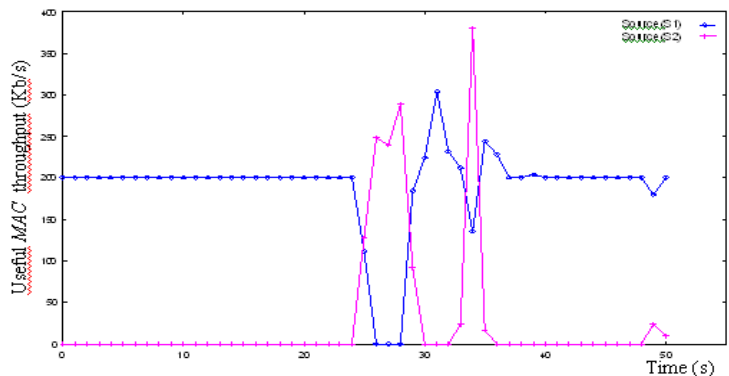

Figure 19. AODV

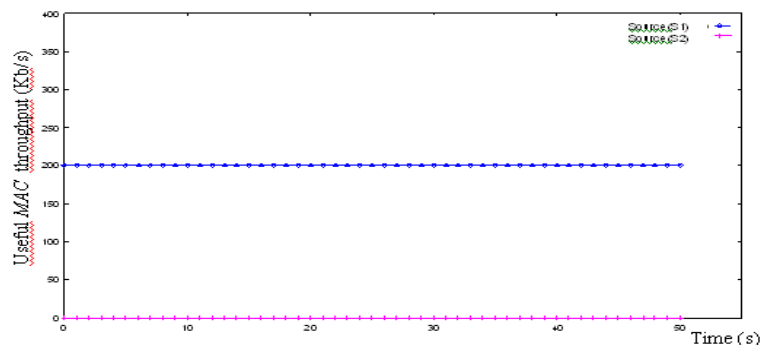

Figure 20. AODVwithQoS

For the routing protocol AODV (Figure 19), no CAC mechanism or resource reservation is used. For it the second flow comes to be added to the others and they shared the medium. This is translated by instability of the useful MAC throughput at the level of both sources of data. The CAC mechanism joined to that resource reservation is speeded up in our protocol AODVwithQoS. The acquired useful MAC throughput is introduced in (Figure 20): In fact, the first data flow of source (S1) issues a RREQ to (D). Nodes, receiving this RREQ, perform CAC and since available bandwidth satisfy the requested bandwidth, then (D) reserve asked resource 200Kb/s in favour of (S1). That issues the flow without any degradation. At t=25seconds, the second source (S2) plans issue its flow with throughput 200Kb/s also and initializes RREQ in this view. And the residual bandwidth is deficient and cannot satisfy (S2) .La RREQ is rejected and no path is established between (S2) and (D) who will not be able to communicate until the end of simulation.

By measuring PDR, we recovered 85 % by applying AODV against 100 % by applying AODVwithQoS. The 15 % of loss for AODV have the saturation of the channel as reason what can cause congestion at intermediate nodes. Let us signal that 15 % of loss is of a rather critical importance for a routing protocol demanding QoS.

### 6.2   Scenario 2 : network  with 50 mobile nodes

Scenario is formed of 50 mobile nodes in movement according to the model RWM. Four sources of data from with the throughput is 100Kb/s. Simulation lasts 60seconds. The measure of useful MAC throughput at every source is introduced by (Figure 21,22).

235

International Journal of Computer Networks & Communications (IJCNC) Vol.3, No.1, January 2011

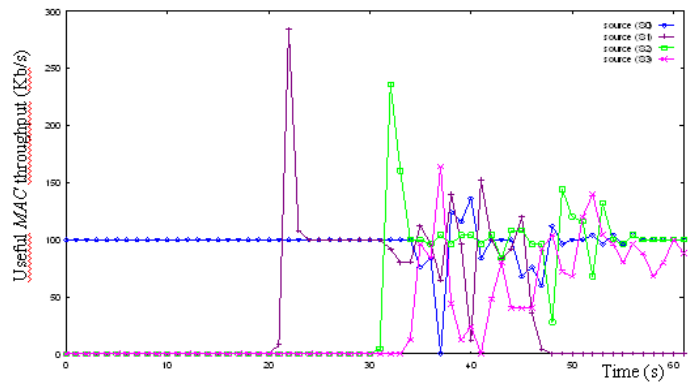

Figure 21. AODV

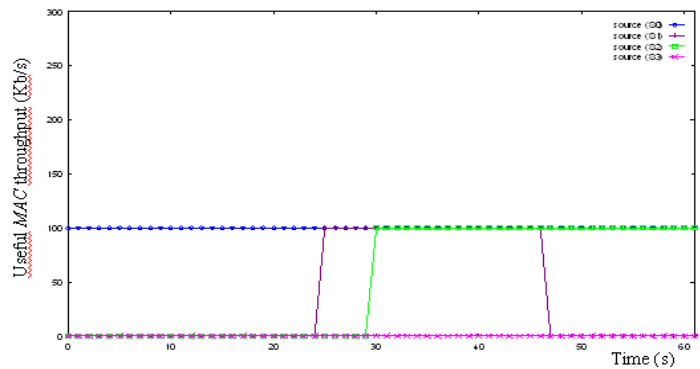

Figure 22. AODVwithQoS

The first source, having asked 100Kb/s, reserves bandwidth and emit its data packets with the requested bandwidth (Figure 22). The same thing is noted for the second and the third source. None of the granted throughput is degraded throughout simulation. However, the fourth source fails in his CAC and cannot reserve 100 Kb/s requested for the emission of its traffic. Our mechanism is efficient as regards CAC and the conservation of the reserved bandwidth. QoS is guaranteed and preserved.
The useful MAC throughput recorded during the simulation of AODV (Figure 21) show however a variation of the throughput of sources. In fact, in t=30 seconds, the third flow comes to add up in the two other previous flows, this flow is going to degrade the throughput of the second source. The third source is going to occupy throughput superior to 200K/s at the expense of the second source's throughput which degrades.

### 6.3 Scenario 3: Valuation of the cost of our mechanism

To assess the cost of our routing mechanism, we offer to count the number of routing packets (Overhead) (11). In fact, this metric measures the number of routing packets transmitted by the different mobile nodes of network.

$$overhead = \sum (RREQ + RREP + RERR) \quad (11)$$

We consider two sources of CBR flow (200Kb/s). The number of nodes, forming simulated network, varies 10, 20, 30, 40 and 50 nodes. The size of packets is 500 bytes and simulation is of 100seconds. (Figure 27)

236

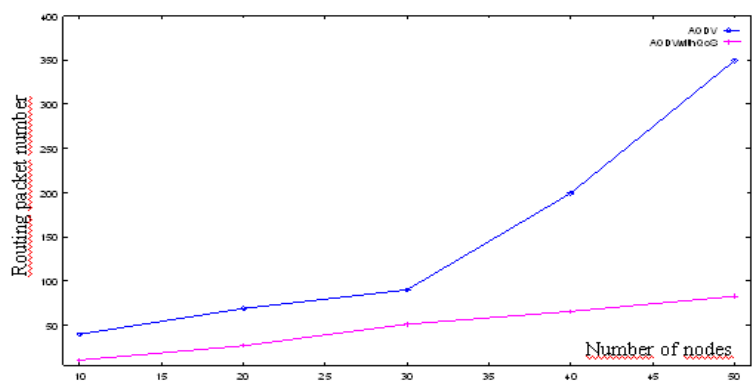

Figure 27. Overload versus nodes number

The curve introduces the necessary total number of control message to establish and maintain paths according to the number of nodes. AODVwithQoS generates fewer control messages than AODV because the CAC eliminates paths not having enough resources in term of residual bandwidth. The elimination of these paths leads to avoid the inundation of network in vain by the research packets of path. Furthermore when network becomes congested, AODV tries to rebuild paths by generating a gust of control packets.

## 7. CONCLUSION

This paper is interested to a critical point on Ad hoc mobile networks: the QoS routing. To assure a routing QoS, we offered a mechanism of determination of the occupation factor of the medium radio. In order to do that, we studied the possible metrics that can influence the estimate of the residual bandwidth to know the phenomenon of interference, RTS/CTS exchanges etc. Our technique exploits the fact that a node can estimate the channel occupancy by monitoring its environment. The aim of available bandwidth estimation is to serve as a basis for admission control of flows sharing the network. In fact, routing strategy is based on admission control for data flows. This admission control is joined to a mechanism of reservation bandwidth during the answer of path if ever the flow being discussed is accepted.

We provide a non-intrusive estimation meaning that it does not generate any additional traffic to perform the evaluation. We showed by simulations that our technique provides an accurate estimation of the available bandwidth on wireless links in many ad hoc configurations. We also showed that PDR assured by AODVwithQoS is better than that of AODV.

International Journal of Computer Networks & Communications (IJCNC) Vol.3, No.1, January 2011

**Authors**

**Heni KAANICHE** received the Engineering Degree from the National School of Engineering of Sfax, Tunisia, in July 2001 and he completed his master in the Network and Performance team of the LIP6 laboratory. He received his Master Degree in Computer Networks and Telecommunications from the University Pierre and Marie Curie (Paris VI), France, in September 2002. He is a researcher in the CRISTAL laboratory at the National School of Computer Science, ENSI, Manouba, Tunisia. His main research concerns Wireless networks, Mobility and Ad Hoc networks. 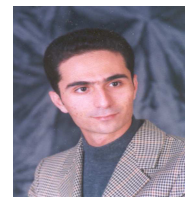

**Fatma LOUATI** (born 1982) received the Engineering Degree from the National School of Engineers of Sfax (ENIS-Tunisia) June 2006. She received his Master Degree in Computer Information Systems and New Technologies from the Faculty of Economics and Management of Sfax (FSEG-TUNISIA) in coordination with the MEDIATRON Unit at the Higher School of Communication of Tunis (SUP'COM-TUNISIA) October 2008. 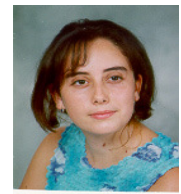

**Mounir FRIKHA** is an Associate Professor at the high School of Communication in Tunis (SUPCOM). He has his PhD in 1999 at the Technical University of Braunschweig in Germany. From 1999 to 2001 he was a project leader in GSM planning in Siemens and he is from 2001 to now an Associate Professor at SUPCOM. His research Areas are Mobility in IP networks, Wireless networks and AD HOC networks.

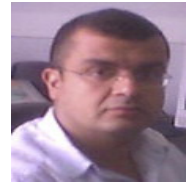

**Farouk KAMOUN,** PhD. (born 1946) is a Tunisian computer scientist and professor of computer science at ENSI (Ecole Nationale Sciences Informatiques : Computer Science School of The University of Manouba, Tunisia). He contributed in the late 1970s to significant research in the field of computer networking in relation with the first ARPANET network. He is also one of the pionniers of the development of the Internet in Tunisia in the early 1990s.

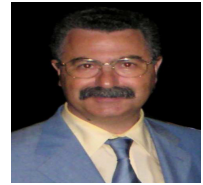